%% file: sparsecount_main_0413.tex
\documentclass[11pt]{article}
\usepackage{times}
\usepackage{alltt}
\usepackage{ulem}
\usepackage{amsfonts}
\usepackage{amssymb}
\usepackage{graphicx}
\usepackage{setspace}
\usepackage{amsmath}
\usepackage{amsthm}
\usepackage{palatino}
\usepackage{todonotes}
\usepackage{paralist}
\usepackage{epstopdf}
\usepackage[top=8pc,bottom=8pc,left=8pc,right=8pc]{geometry}
\usepackage{algorithm}
\usepackage{algpseudocode}
\usepackage{subcaption}
\usepackage{bm}
\usepackage[titletoc,title]{appendix}

\usepackage[round,sort,comma]{natbib}
\include{math-commands}

\graphicspath{{../../figures/}{../figures/}{./figures/}{./}}

\def\T{{ \mathrm{\scriptscriptstyle T} }}

\usepackage{color}
\usepackage{hyperref}
\definecolor{darkblue}{rgb}{0.0,0.0,0.3}
\hypersetup{colorlinks,breaklinks,linkcolor=darkblue,urlcolor=darkblue,
            anchorcolor=darkblue,citecolor=darkblue}
\title{Inference on High-Dimensional Sparse Count Data}

\author{Jyotishka Datta  \footnote{Department of Statistical Science, Duke University email:jd298@stat.duke.edu} 
\and David B. Dunson \footnote{Department of Statistical Science, Duke University, email: dunson@duke.edu}}
\date{\vspace{-0.1cm} \today}
\IfFileExists{upquote.sty}{\usepackage{upquote}}{}

\begin{document}
\maketitle

\begin{abstract}
In a variety of application areas, there is a growing interest in analyzing high dimensional sparse count data, with sparsity exhibited by an over-abundance of zeros and small non-zero counts.  Existing approaches for analyzing multivariate count data via Poisson or negative binomial log-linear hierarchical models with zero-inflation cannot
flexibly adapt to the level and nature of sparsity in the data.  We develop a new class of continuous local-global shrinkage priors tailored for sparse counts.  Theoretical properties are assessed, including posterior concentration, stronger control on false discoveries in multiple testing, robustness in posterior mean and super-efficiency in estimating the sampling density. Simulation studies illustrate excellent small sample properties relative to competitors. We apply the method to detect rare mutational hotspots in exome sequencing data and to identify cities most impacted by terrorism.
\end{abstract}


\section{Introduction}

This paper considers the problem of modeling of high-dimensional sparse count data $y = (y_1,\ldots, y_n)^{\T}$ having an excess of values near zero.  In many applications, the rates of event occurrence are very small at an over-whelming majority of `locations', with substantially higher rates at a subset of locations.  For example, $y_i$ may correspond to the number of mutations observed at location $i$ in the genome or the number of terrorist activities in city $i$.  There is a rich literature on modeling of count data focused on accommodating over-dispersion relative to the Poisson distribution and zero-inflation in the counts.  Such models are insufficiently flexible to accommodate the common setting in which the data have an abundance of very small but not necessarily zero counts.  This article proposes a novel shrinkage prior for the rate parameters, which simultaneously accommodates highly sparse and heavy tailed signals $\theta$, while maintaining computational tractability and theoretical guarantees. The proposed shrinkage prior is built upon the Gauss Hypergeometric distribution \citep{armero1994prior} proposed for modeling the traffic intensity of an $M/M/1$ queue in equilibrium.

For simplicity, we focus on the model $y_i \sim \mbox{Poi}(\theta_i)$, independently for $i=1,\ldots,n$ with $\theta = (\theta_1, \ldots, \theta_n)^{\T}$.  However, our proposed approach can be used directly for more elaborate models that let $y_i \sim \mbox{Poi}(\theta_i \eta_i)$, with $\theta_i$ a random effect and $\eta_i$ a structured part characterizing dependence on covariates, hierarchical designs, spatial or temporal structure and other complications.  In the simple case, estimating the $\theta$ vector is commonly referred to as the Poisson compound decision problem.  Empirical Bayes approaches have been popular in this context, dating back to  Robbins' formula \citep{robbins1956empirical}. 
In Robbins' approach, the $\theta_i$'s are assumed to be independent draws from a distribution $G(\cdot)$, and the goal is to estimate $\theta = (\theta_1, \ldots, \theta_n)^{\T}$ depending on the observations $y = (y_1, \ldots, y_n)^{\T}$. The accuracy of an estimator $\hat{\theta} = \delta(y) = \{\delta_1(y_1), \ldots, \delta_n(y_n)\}^{\T}$ is assessed by the risk $W(\delta) = E_{\theta} \vectornorm{\delta(y)-\theta}^2$. The Bayes estimator that minimizes $W(\delta)$ assumes a simple form for the Poisson kernel as 
\beq
\delta_i(y_i)=\frac{ \int \theta_i p(y_i \mid \theta_i) dG(\theta_i)}{\int p(y_i \mid\theta_i) dG(\theta_i)} = \frac{(y_i+1)P_Y(y_i+1)}{P_Y(y_i)}, \nonumber
\eeq
where $P_Y({\cdot})$ is the marginal distribution of $Y$. Robbin's frequency ratio estimator uses the empirical frequencies $\hat{P}_Y(y) = n^{-1} \sum_{i=1}^{n} I(y_i = y)$ to estimate $P_Y$.  Performance is deteriorated by slow convergence of $\hat{P}_Y \to P_Y$ \citep{brown2013poisson}. \cite{brown2013poisson} proposed a three-stage smoothing adjustment that shows substantial improvement in total Bayes risk $nW(\delta)$ in simulation studies. \cite{koenker2014convex} proposed a more efficient approach based on nonparametric maximum likelihood estimation \citep{kiefer1956consistency} that estimates $\theta$ by maximizing the likelihood with respect to the unknown distribution $G$. 

This article develops a hierarchical Bayesian model that allows for sparsity in $G$, while maintaining the ability to capture large signals.  The proposed prior is adaptive to the degree of sparsity in the data, and is inspired by local-global shrinkage priors for sparse Gaussian means and linear regression \citep{carvalho2010horseshoe,armagan2011generalized,armagan2013generalized,bhattacharya2014dirichlet}.  Such priors are structured as scale mixtures of Gaussians for convenience in computation, with corresponding theory support when the true mean or regression vector is mostly zero \citep{bhattacharya2014dirichlet,van2014horseshoe}.  Naively, one could apply such priors to the coefficients in Poisson log-linear models, but such formulations lack the computational advantages obtained in the Gaussian case and fail to represent sparsity corresponding to rates close to zero.  

Our proposed model induces inflation of small counts in a continuous manner, which has important advantages over zero-inflated Poisson models and their many variants, such as the zero-inflated generalized Poisson and zero-inflated negative binomial \citep{yang2009testing}.  Examples such as \S \ref{sec:wgs} and \S \ref{sec:gtd} produce an abundance of `low counts'. Under a zero-inflated model, the $\theta_i$'s are set to zero with probability $p$ or sampled from a simple parametric distribution, typically either a degenerate distribution at a single $\theta$ value or a gamma distribution.  This restrictive parametric form limits performance, as we will illustrate.  The two component mixture form also leads to computational instability in high-dimensional sparse count examples due to high dependence between the parameters $p$ and $\theta$. 

\section{Shrinkage priors for count data}\label{sec:gl}

If $\theta_i \sim \mbox{Ga}(\alpha,\beta_i)$, the marginal distribution of $y_i$ is negative binomial with variance, $\alpha{\beta_i}^{-1}(1+\beta_i^{-1})$, higher than the mean, $\alpha{\beta_i}^{-1}$.  To allow zero-inflation, the usual approach would mix a Poisson or negative binomial distribution with a degenerate distribution at zero.  We instead choose a prior for $\theta_i$ with a pole at zero leading to a spike in the marginal distribution for $y_i$ at zero. Our Poisson-Gamma hierarchical model can be expressed as: 
$$y_i \sim \mbox{Poi} (\theta_i),\quad
\theta_i \sim \mathrm{Ga}(\alpha, \lambda_i^2 \tau^2),\quad
\lambda_i \sim p(\lambda_i^2),\quad \tau \sim p(\tau^2),
$$
where $p(\lambda_i^2)$ and $p(\tau^2)$ are densities for $\lambda_i^2$ and $\tau^2$, respectively.  Marginalizing out $\theta_i$ and writing $\kappa_i = 1/(1+\lambda_i^2\tau^2)$, the model can be written as: 
\begin{align}
p(y_i \mid \lambda_i, \tau) & \propto \left(\frac{\lambda_i^2\tau^2}{1+ \lambda_i^2\tau^2} \right)^{y_i}  \left( \frac{1}{1+ \lambda_i^2\tau^2} \right)^\alpha, \label{eq:hier1} \\
p(y_i \mid \kappa_i) & \propto (1-\kappa_i)^{y_i} \kappa_i^\alpha \Rightarrow [y_i \mid \kappa_i] \sim \mathrm{NB}(\alpha, 1-\kappa_i). \label{eq:hier2}
\end{align}
In other words, given $\kappa_i$, $y_i$ follows a negative binomial distribution with size $\alpha$ and probability of `success' $1-\kappa_i$. The posterior distribution and mean of $\theta_i$ given $y_i$ and $\kappa_i$ are respectively 
\beq
p(\theta_i \mid y_i, \kappa_i) \sim \mathrm{Ga}(y_i + \alpha, 1-\kappa_i),\quad E (\theta_i \mid y_i, \kappa_i) = (1-\kappa_i)(y_i+\alpha). \nonumber
\eeq
Hence, the parameter $\kappa_i$ can be interpreted as a random shrinkage factor pulling the posterior mean towards $0$. 

Priors on shrinkage factors having a $U$-shaped distribution are appealing in shrinking small signals to zero while avoiding shrinkage of larger signals.  In normal linear models, such priors have been widely used and include the horseshoe \citep{carvalho2010horseshoe}, generalized double Pareto \citep{armagan2013generalized}, three-parameter beta \citep{armagan2011generalized} and Dirichlet-Laplace \citep{bhattacharya2014dirichlet}.  In our motivating sparse count applications, we require additional flexibility in the mass of the shrinkage parameter $\kappa_i$ around $0$ and $1$ due to the occurrences of very low counts in addition to zeros.  For example, small counts can arise due to measurement errors and should be separated from true rare events to the extent possible.  This requires careful treatment of the prior for $\kappa_i$.

Consider the three-parameter beta prior \citep{armagan2011generalized}:
\begin{equation}
p(\kappa_i \mid a, b, z) \propto (1-\kappa_i)^{a-1} \kappa_i^{b-1} (1+ z \kappa_i)^{-(a+b)}\; \mbox{ for } 0 < \kappa_i, z < 1, a, b>0. \label{eq:tpb}
\end{equation}
\cite{armagan2011generalized} recommend $a,b \in (0,1)$ leading to both Cauchy-like tails and a kink at zero. For example, $a = b = 1/2$ and $z = 0$ in \eqref{eq:tpb} leads to $\kappa_i \sim \mathrm{Be}(1/2,1/2)$, where $\mathrm{Be}(a,b)$ denotes the beta distribution with parameters $a$ and $b$. The `horseshoe'-shaped $\mathrm{Be}(1/2,1/2)$ prior combined with the likelihood in \eqref{eq:hier2} produces a $(1-\kappa_i)^{y_i-1/2}$ term in the posterior, which leaves all non-zero $y_i$'s unshrunk. For $a=b=1/2$ and $z \to -1$, the corresponding term in the posterior would be $(1-\kappa_i)^{y_i-3/2}$, which shrinks $y_i \le 1$. 

To extend the flexibility of the prior on $\kappa_i$, while retaining the heavy-tailed property of the induced marginal prior on $\theta_i$, we make the exponent in the final term in \eqref{eq:tpb} a general non-negative parameter $\gamma$. Higher values of $\gamma$ imply shrinkage of larger observations in the posterior. The proposed prior on $\kappa_i$ has the following functional form: 
\beq
\mbox{\small GH}(\kappa_i \mid a,b, z, \gamma) = C \kappa_i^{a-1} (1-\kappa_i)^{b-1}(1+z \kappa_i)^{-\gamma} \mbox{ for } \kappa_i \in (0,1),  \label{eq:GH}
\eeq
where $C^{-1} = \beta(a,b) _2F_1(\gamma,a,a+b,-z)$ is the norming constant, with $\beta(a,b) = \Gamma(a)\Gamma(b){\Gamma(a+b)}^{-1}$ and $_2F_1$ the Gauss Hypergeometric function defined as:
\begin{equation*}
_2F_1(a,b,c,z) = \sum_{k=0}^{\infty} \frac{(a)_k (b)_k}{(c)_k} \frac{z^k}{k!} \mbox{ for } \mid z \mid < 1, 
\end{equation*}
where $(q)_k$ denotes the rising Pochhamer symbol, defined as $(q)_k = q (q+1) \ldots (q+k-1)$ for $k>0$ and $(q)_0 = 1$. Expression (\ref{eq:GH}) corresponds to the Gauss Hypergeometric ({\small GH}) distribution introduced in \cite{armero1994prior} for a specific queuing application. 

\begin{figure}[!ht]
\centering
\begin{subfigure}{0.45\textwidth}
  \includegraphics[width=\linewidth]{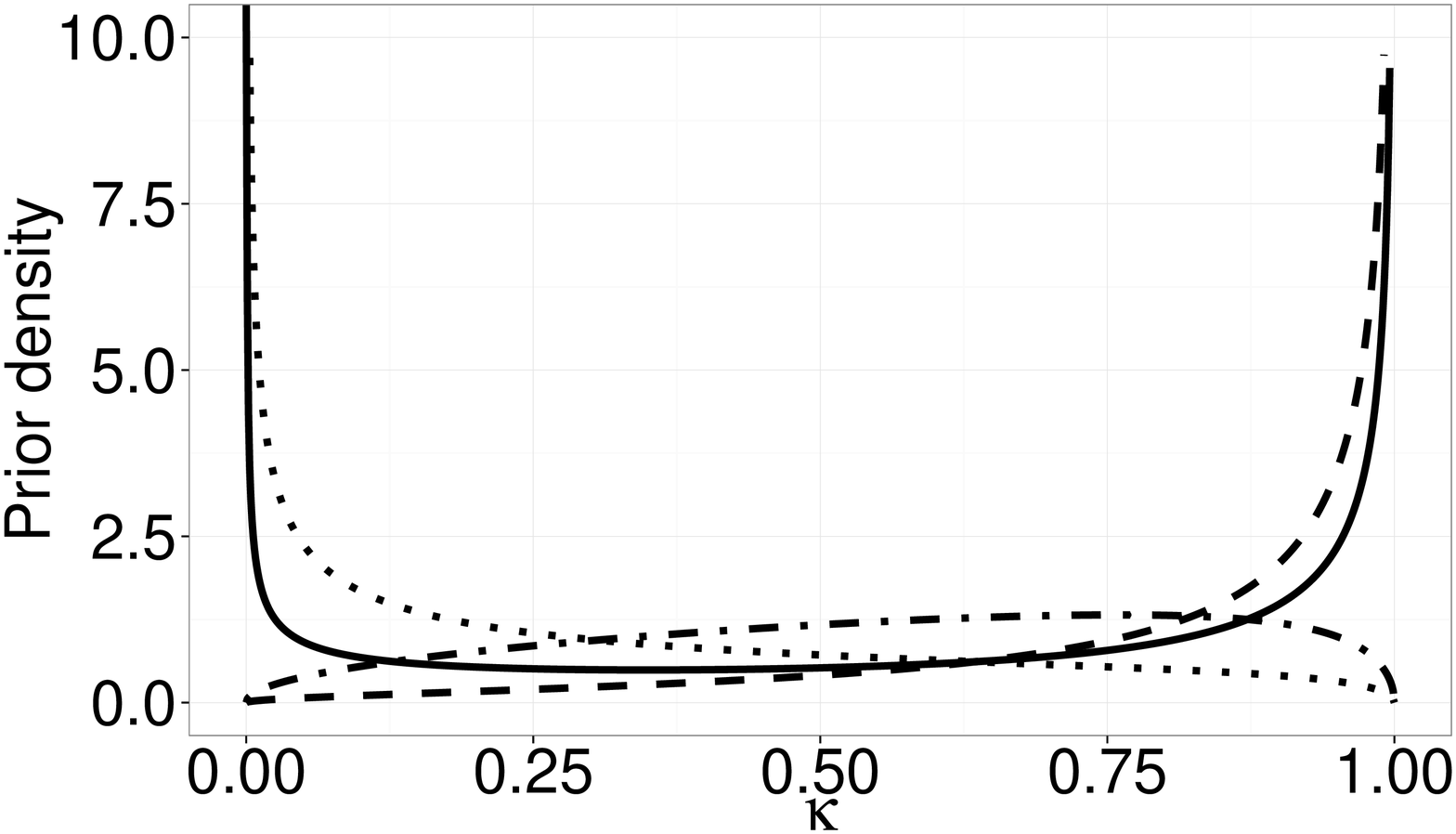}
  \caption{Effect of $a$ and $b$ on the prior density of $\kappa_i$.}
  \label{fig:effect-a}
\end{subfigure}
\hspace{0.1in}
\begin{subfigure}{0.45\textwidth}
  \includegraphics[width=\linewidth]{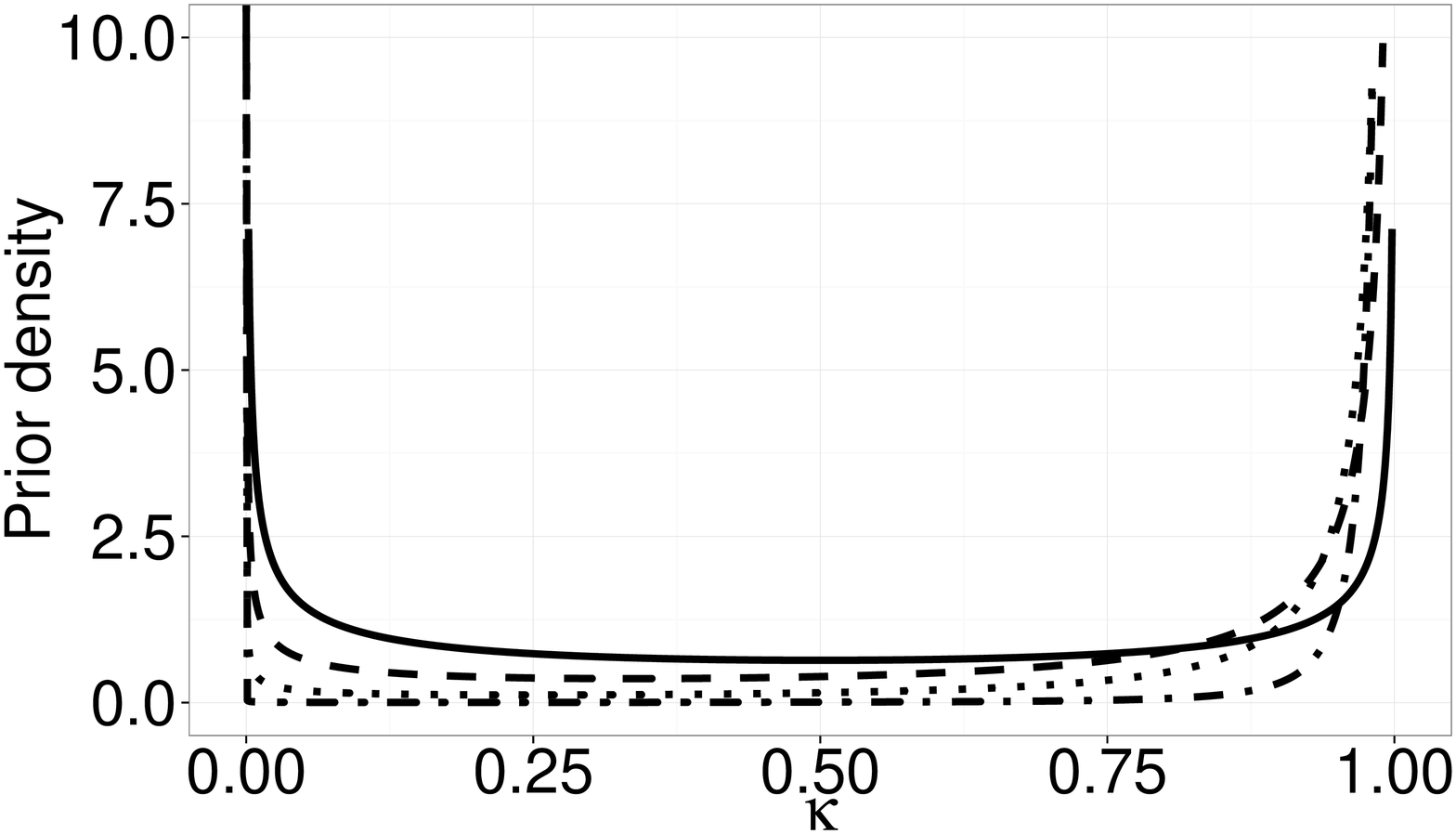}
  \caption{Effect of $\gamma$ on the prior density of $\kappa_i$.}
  \label{fig:effect-gamma}
\end{subfigure}
\caption{A comparison of density of $\kappa_i$ under the Gauss Hypergeometric prior for different values of the hyper-parameters: (a) $a=b=0{\cdot}5$ (solid), $a=0{\cdot}5,b=1{\cdot}5$ (dashed), $a=1{\cdot}5,b=0{\cdot}5$ (dotted), and $a=b=1{\cdot}5$ (dot-dash), (b) $\gamma=0$ (solid), $\gamma=0{\cdot}5$ (dashed), $\gamma=1$ (dotted), and $\gamma=2$ (dot-dash).}
\label{fig:effect}
\end{figure}

Our sparsity-inducing prior on $\kappa_i$ is $\mbox{\small GH}(a = 1/2,b = 1/2, z= \tau^2-1, \gamma)$, where $\tau^2$ is a global shrinkage parameter adjusting to the level of sparsity in the data. The {\small GH} prior is conjugate to the negative binomial likelihood in \eqref{eq:hier2} and will lead to the following posterior distribution: 
\begin{align}
p(\kappa_i \mid y_i, \tau^2, \gamma) & =  \frac{\kappa_i^{\alpha-1/2} (1-\kappa_i)^{y_i-1/2}\{1-(1-\tau^2)\kappa_i\}^{-\gamma}}{\beta(\alpha+1/2,y_i+1/2) _2F_1(\gamma,\alpha+1/2,y_i+\alpha+1,1-\tau^2)} \nonumber \\
\Rightarrow \kappa_i \mid y_i, \tau^2, \gamma & \sim \mbox{{\small GH}}\big(\alpha+1/2, y_i+1/2, \gamma, \tau^2-1 \big). \label{eq:posterior}
\end{align}
Plots of the prior density for different values of the hyper-parameters $a$, $b$ and $\gamma$ for $\tau^2 = 0{\cdot}01$ are given in Figure \ref{fig:effect}. The first column shows the effect of different choices of the parameters $a,b$ when $\gamma = 1/2$ and the second column shows the effect of $\gamma$ on $p(\kappa_i)$ for $a = b= 1/2$. 

As shown in Figure \ref{fig:effect}, the {\small GH} prior results in a U-shaped prior density of $\kappa_i$ for $a = b = 1/2$ for different values of $\gamma$ and $\tau^2$. This is a general class that includes the horseshoe prior. Putting a standard half-Cauchy prior, denoted as $C^{+}(0,1)$, on $\lambda_i$ in the hierarchical set-up \eqref{eq:hier1}-\eqref{eq:hier2} leads to the same posterior as under the prior $\mbox{\small GH}(1/2,1/2,\gamma = 1, 1-\tau^2)$. The half-Cauchy $C^{+}(0,1)$ prior introduced by \cite{gelman2006prior} is a default shrinkage prior that leads to the Horseshoe prior for the sparse normal means problem \citep{carvalho2010horseshoe}. However, the {\small GH} prior leads to lower estimation and misclassification error relative to the horseshoe prior for sparse counts, as we show in the numerical study section \S\ref{sec:app}. 

The $k^{th}$ posterior moment for $\kappa_i$ given $y_i$, $\tau$, and $\gamma$ can be written as 
\beq
E (\kappa_i^k \mid y_i, \tau, \gamma) = \frac{\beta(k+\alpha+1/2,y_i+1/2) _2F_1(\gamma,k+\alpha+1/2,y_i+\alpha+1+k,1-\tau^2)}{\beta(\alpha+1/2,y_i+1/2) _2F_1(\gamma,\alpha+1/2,y_i+\alpha+1,1-\tau^2)}. \label{eq:post-moment}
\eeq
Equation \eqref{eq:post-moment} provides a way to rapidly calculate an empirical Bayes estimate of posterior mean $E (\kappa_i \mid y_i, \hat{\tau}, \hat{\gamma})$. We will show in \S \ref{sec:theory} that the posterior distribution of $\kappa_i$ for the {\small GH} prior will concentrate near $0$ or $1$ for `large' or `small' observations, respectively. 

\begin{remark}
The Gauss-Hypergeometric distribution in \cite{armero1994prior} was introduced in a completely different context of modeling a one-dimensional parameter $\rho$, the traffic intensity in a $M/M/1$ queue. We, on the other hand, use the {\small GH} prior on each shrinkage coefficient for modeling the multivariate, sparse mean vector $\theta$, where the goal is to separate the smaller parameters from the larger ones. 
\end{remark}

\subsection{Properties of the prior}\label{sec:priorprop}

We use the following lemma to derive the marginal prior on $\theta_i$ for $\tau^2 =1$. The proofs are deferred to the Appendix.  
\begin{lemma}\label{lemma:1}
Let $Z \sim \mbox{Ga}(\alpha,\beta)$, then $E\{1/(c+Z)\} = \beta^{\alpha} c^{\alpha-1} e^{\beta c} \Gamma(1-\alpha, \beta c)$, where $\Gamma(s,z)$ is the upper incomplete gamma function defined as $\int_{z}^{\infty} t^{s-1}\mathrm{e}^{t-1} dt$.
\end{lemma}
\begin{proposition}\label{prop:marginal}
Let $\tau^2 = 1$. For $\theta \sim \mathrm{Ga}(\alpha, 1/\lambda^2)$ and $\lambda \sim C^+(0,1)$, the prior density on $\theta_i$ can be written as: 
\beq
p(\theta_i) = \frac{2}{\pi \Gamma(\alpha) } \int_{0}^{\infty} {\theta_i}^{\alpha-1} \mathrm{e}^{-\theta_i/\lambda^2}   \frac{{\lambda}^{-2\alpha}}{\lambda^2+1} d\lambda = \frac{1}{\sqrt{\pi} \beta(1/2,\alpha)}  \mathrm{e}^{\theta_i} {\theta_i}^{\alpha-1} \Gamma \left(1/2 - \alpha,\theta_i \right),  \label{eq:prop1}
\eeq
where $\Gamma(s,z)$ is the upper incomplete gamma function. 
\end{proposition}
An upshot of Proposition \ref{prop:marginal} is that we can use sharp bounds for the upper incomplete Gamma function $\Gamma(s,z)$ to derive useful bounds for the marginal density of the prior on $\theta_i$. As we show in Corollary \ref{cor:marginal} and its proof in Appendix 1, the bounds on $\Gamma(1/2-\alpha, \theta_i)$ determine the bounds on $p(\theta_i)$, and play a crucial role in producing a spike in $p(\theta_i)$ near zero and a polynomially heavy tail. The plot of the prior density $ p(\theta_i \mid \alpha = 1)$ as well as $ p(\theta_i \mid \alpha = 1/2)$ near the origin and the tails are given below, along with a few other candidate prior densities.  

\begin{figure}[!ht]
\centering
\begin{subfigure}{0.45\textwidth}
  \centering
  \includegraphics[width=\linewidth]{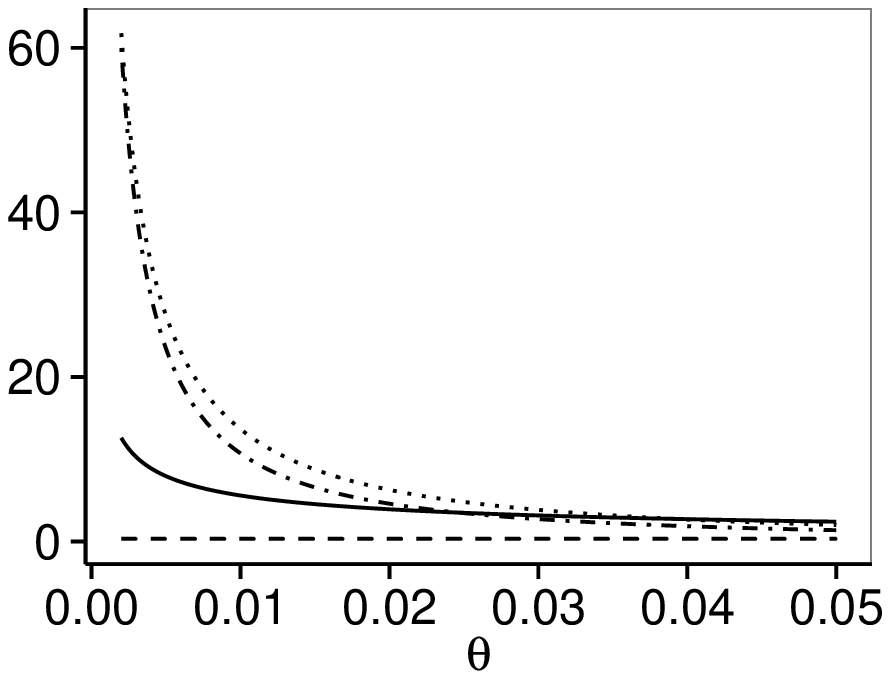}
  \caption{Prior densities near origin}
  \label{fig:prior-origin}
\end{subfigure}
\begin{subfigure}{0.45\textwidth}
 \centering
  \includegraphics[width=\linewidth]{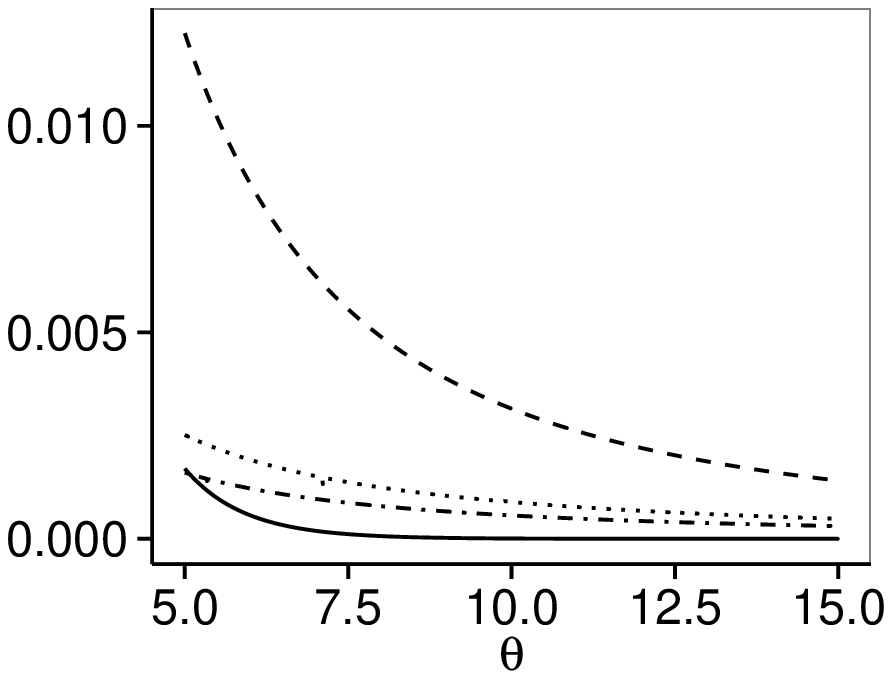}
  \caption{Tails of prior densities}
  \label{fig:prior-tails}
\end{subfigure}
\caption{Comparison of density of $\theta_i$ under different priors: Gamma (solid), Cauchy (dashed), Gauss-Hypermetric with $\alpha=1/2$ (dotted), and Gauss Hypergeometric with $\alpha=1$ (dot-dash).} 
\label{fig:prior}
\end{figure}

\begin{Cor}\label{cor:marginal}
The prior density on $\theta_i$ satisfies the following sharp upper and lower bounds for $\alpha = 1$ and $\alpha = 1/2$ respectively: 
\begin{align}
\frac{1}{\pi} \left( \frac{1}{\sqrt{\theta_i}} - \frac{2}{\sqrt{\theta_i}+\sqrt{\theta_i+4/\pi}} \right) & \leq p \left(\theta_i \mid \alpha = 1 \right) \leq \frac{1}{\pi}\left( \frac{1}{\sqrt{\theta_i}} - \frac{2}{\sqrt{\theta_i}+\sqrt{\theta_i+2}} \right) \nonumber \\ 
\frac{1}{2\pi^{3/2} } \theta_i^{-1/2} \log \left( 1+ \frac{2}{\theta_i} \right) & \leq  p \left(\theta_i \mid \alpha = 1/2 \right) \leq \frac{1}{\pi^{3/2} } \theta_i^{-1/2} \log \left(1+\frac{1}{\theta_i} \right)  \label{eq:bounds2}
\end{align}
\end{Cor}

\subsection{Impact of hyper-parameters}\label{sec:prop}

\begin{figure}[!ht]%
\includegraphics[width=\linewidth]{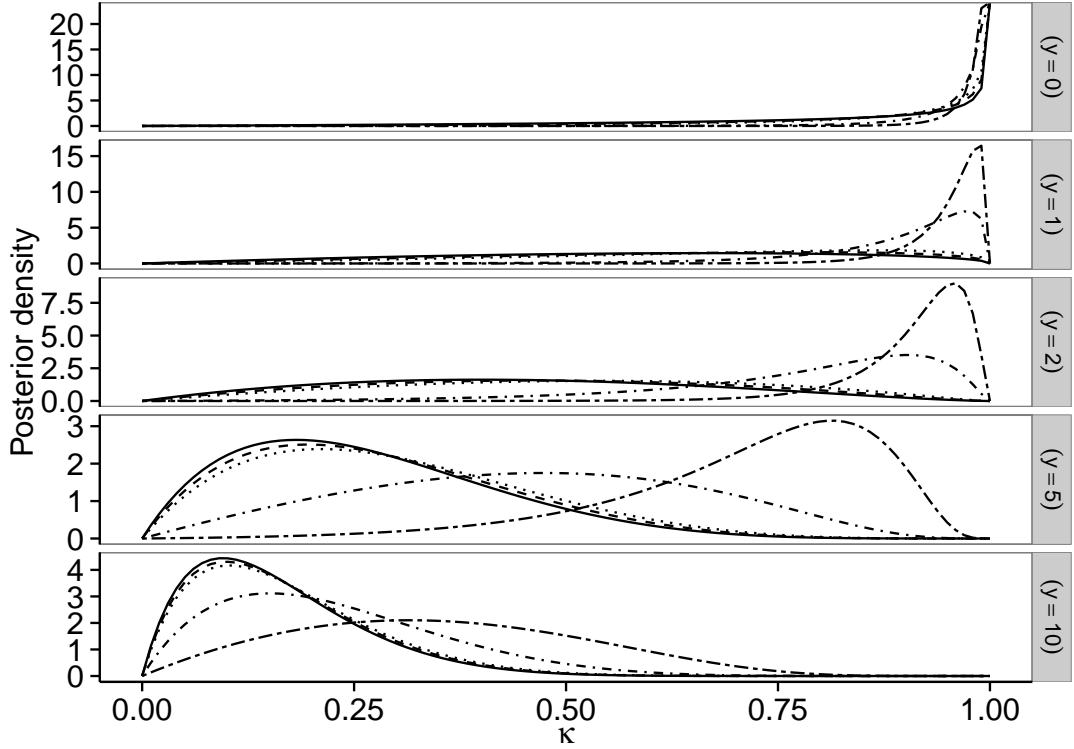}%
\caption{Posterior distribution of the shrinkage parameter $\kappa_i$ under the Gauss Hypergeometric prior for different values of the hyper-parameter $\gamma$: $\gamma=0$ (solid), $\gamma=0{\cdot}5$ (dashed), $\gamma=1$ (dotted), $\gamma=5$ (dot-dash), $\gamma=10$ (two-dash). Each row corresponds to a different value of $y_i \in \{0, 1, 2, 5, 10\}$.}
\label{fig:post}%
\end{figure}

The Gauss Hypergeometric posterior distribution in \eqref{eq:posterior} depends on three hyper-parameters  $\alpha$, $\gamma$ and $\tau$. These hyper-parameters are chosen to ensure that the posterior drives most of its mass to $\kappa_i = 0$ and $\kappa_i = 1$ for appropriate values of $y_i$. The parameter $\alpha$ controls the shape of the marginal prior on $\theta_i$ as shown in Corollary \ref{cor:marginal} and Figure \ref{fig:prior} in \S \ref{sec:priorprop}, with smaller values, such as $\alpha = 1/2$, leading to heavier tails and a higher spike near zero. The value of $\alpha=1/2$ is recommended as a default, leading to super-efficiency in Kullback-Leibler risk as shown in the Supplement. The parameter $\gamma$ controls shrinkage adapting to the abundance of small counts in the data. Figure \ref{fig:post} shows the combined effect of $\gamma$ and $y_i$ on the posterior distribution of $\kappa_i$. For a small value of $\gamma$, the posterior distribution given a large $y_i$, such as $y_i = 10$, concentrates near $\kappa_i=0$, indicating no shrinkage. However, a small value of $y_i$, such as $y_i = 1$, reinforces the shrinkage by $\tau$ and the posterior concentrates near $\kappa_i=1$, leading to strong shrinkage of low counts. A direct outcome of the extra flexibility is a stronger control of the type I  error probability, as shown in Theorem \ref{th:type-1} in \S \ref{sec:t1} and simulation studies in supplementary \S S.4. These clearly show the role of the `thresholding' parameter $\gamma$ in controlling the shape of the posterior distribution. Finally, the parameter $\tau$ controls global shrinkage as well as the rates of posterior concentration. 

\section{Theoretical Properties}\label{sec:theory}

In this section, we study theoretical properties, with the main results showing adaptability to different types of sparsity through flexible posterior concentration manifesting in a stronger control on false discoveries in multiple testing and robustness to large signals. We first formalize the behavior of the posterior distribution of the shrinkage weight $\kappa_i$ under the {\small GH} prior. Towards this, we provide bounds for posterior mass in neighborhoods of $0$ and $1$ depending on the magnitude of the observation $y_i$, the `thresholding' parameter $\gamma$ and the global shrinkage parameter $\tau$.  These results imply that thresholding $\kappa_i$ to produce a multiple testing rule leads to a type I error rate that decreases with $\gamma$ when the true $\theta_i$s are independent draws from a discrete mixture of gamma distributions.  We additionally establish robustness properties of the posterior mean, showing that $| E(\theta_i \mid Y_i = y_i) - y_i |$ is not too large. As a supplemental result, we show that the posterior mean density estimator under the proposed prior has a faster rate of convergence in terms of Ces\'aro risk, than any prior with a bounded mass near zero, when the true parameter is zero. 

\subsection{Flexible Posterior Concentration} 
Since the posterior mean for $\theta_i$ can be written as $(1-\hat{\kappa_i}) (y_i +\alpha)$, it is natural to expect that the posterior distribution of $\kappa_i$ puts increasing mass at zero as $y_i$ becomes large relative to the hyper-parameter $\gamma$.  On the other hand, the posterior mass of $\kappa_i$ concentrates near $1$ for values of $y_i$ small relative to $\gamma$. Indeed, the plots in Figure \ref{fig:post} show the concentration of the posterior distribution at either extreme of the $\kappa_i$ scale depending on the magnitude of $\gamma$. This leads to great flexibility in the shrinkage profile through the posterior mean $\hat{\theta_i} = (1-\hat{\kappa_i})(y_i+\alpha)$ by differential shrinkage depending on the value of $\gamma$. We prove this formally in the two theorems that follow. Theorem \ref{theorem:large-y} shows that the posterior probability of an interval around $1$ approaches $0$ as $y_i$ goes to infinity.  Theorem \ref{theorem:small-tau} shows the posterior distribution of $\kappa_i$ instead concentrates near $1$ when $y_i < \gamma- 1/2$ and $\tau \rightarrow 0$.  

\begin{theorem}\label{theorem:large-y}
Suppose $y_i \sim \mbox{Poi}(\theta_i)$ and let $p(\kappa_i \mid y_i,\tau, \gamma)$ denote the posterior density of $\kappa_i$ given $y_i$ and fixed $\tau$,$\gamma$ for the {\small GH} prior $(\kappa_i \mid \tau, \gamma) \sim \mbox{\small GH}(1/2,1/2,\gamma,\tau^2-1)$. Then the posterior density of $\kappa_i$ satisfies the following: 
\begin{align*}
\mbox{pr}(\kappa_i > \eta \mid y_i,\tau,\gamma) & \leq C(\eta)\frac{1}{1-\tau^2} \left(1 + \frac{\tau^2\eta}{1-\eta} \right)^{-(y_i-1/2-\gamma)} \\
\Rightarrow (\kappa_i \mid y_i, \tau,\gamma) & \to \delta_{\{0\}} \mbox{ as } y_i \to \infty 
\end{align*}
where $\delta_{\{0\}}$ denotes the point mass at zero. 
\end{theorem} 

\begin{theorem}\label{theorem:small-tau}
Suppose $y_i \sim \mbox{Poi}(\theta_i)$ and let $p(\kappa_i \mid y_i, \tau, \gamma)$ denote the posterior density of $\kappa_i$ given $y_i$ and fixed $\tau$, $\gamma$ for the {\small GH} prior $(\kappa_i \mid \tau, \gamma) \sim \mbox{\small GH}(1/2,1/2,\gamma,\tau^2-1)$. Then the posterior density of $\kappa_i$ satisfies the following for all values of $y_i \leq \gamma- 1/2$:  
\begin{align*}
\mbox{pr}(\kappa_i < \eta \mid y_i ,\tau, \gamma) & \leq \left( \frac{\tau^2}{1-\eta} \right)^{d} \\
\Rightarrow (\kappa_i \mid y_i , \tau, \gamma) \to \delta_{\{1\}} \mbox{ as } \tau \to 0 & \mbox{ where } d = (\gamma-1/2-y_i) > 0
\end{align*}
where $\delta_{\{1\}}$ denotes the point mass at one. 
\end{theorem} 
\subsection{Tighter Control on False Discoveries} \label{sec:t1}
The `two-groups' model provides a natural framework for incorporating sparsity, where the $\theta_i$'s are independent and identical draws from a scale mixture of two gamma distributions: 
\begin{equation}
\theta_i \sim (1-p) \mbox{Ga}(\alpha, \beta) + p \mbox{ Ga}(\alpha, \beta+\delta), \mbox{ where } p \in (0,1). \label{eq:twogroups}
\end{equation}
We are interested in testing $H_{0i}: \theta_i \sim \mbox{Ga}(\alpha, \beta)$ against 	$H_{1i}: \theta_i \sim \mbox{ Ga}(\alpha, \beta+\delta)$ for $i = 1, \ldots, n$. We set the shape and scale parameters of the null distribution, $\alpha$ and $\beta$, to small values to ensure higher concentration near zero under the null $H_{0i}$, and $\delta$ to a large value relative to $\beta$ so that the prior becomes more `flat' under $H_{1i}$. 
The posterior mean of $\theta_i$ can be written as: 
\beq
E(\theta_i | Y_i) = \bigg\{ (1-\omega_i)\frac{\beta}{1+\beta} + \omega_i \frac{\beta+\delta}{1+\beta+\delta} \bigg\} (Y_i + \alpha ) \equiv \omega_i^* (Y_i+ \alpha), \label{eq:two-groups}
\eeq
where $\omega_i$ and $\omega_i^*$ denote the posterior inclusion probability, $\mbox{pr}(H_{1i} | Y_i)$, and the observation-specific shrinkage weight respectively. For a fixed $\beta > 0$, if $\delta \to \infty$, $\omega_i^*$ converges to $(\omega_i + \beta)(1+\beta)^{-1}$, which is an increasing function of the posterior inclusion probability $\omega_i$ obtained by redistributing the probability mass in the interval $[\beta(1+\beta)^{-1},1]$. This can be used to construct a multiple testing rule with an appropriate threshold for the shrinkage weight. 
Since $\beta$ is unknown in almost all applications, we cluster the $\omega_i^*$'s into two classes and use the decision boundary as the threshold for $\omega_i^*$. Clearly, if $\beta \approx 0$, $\omega_i^* \to \omega_i$, and the testing rule is to reject $H_{0i}$ if $\omega_i^* > 1/2$. 

The {\small GH} prior directly models the shrinkage weight through the hierarchy: $Y_i \sim \mbox{Poi}(\theta_i)$, $\theta_i \sim \mbox{Ga}(\alpha,\kappa_i^{-1}-1)$ and $\kappa_i \sim \mbox{GH}(1/2,1/2, \tau^2-1, \gamma)$. Since the posterior mean of $\theta_i$ is given by $E(\theta_i | Y_i, \tau, \gamma) = \{ 1- E(\kappa_i | Y_i, \tau, \gamma) \}(Y_i+\alpha)$, a comparison with \eqref{eq:two-groups} suggests that the term $1- E(\kappa_i | Y_i, \tau, \gamma)$ mimics the shrinkage factor $\omega_i^*$, and induces a multiple testing rule: 
$$
\mbox{ Reject } H_{0i} \mbox{ if } 1- E(\kappa_i | Y_i, \tau, \gamma) > \xi \; ; i = 1, \ldots, n, 
$$ 
where $\xi$ is a suitably chosen threshold, calculated by clustering the shrinkage weights into two groups, as described above. The resulting multiple testing rule yields excellent performance in terms of type I  error and misclassification rates in our simulation studies. We now establish that the probability of type I  error for the multiple testing rule induced by the {\small GH} prior decreases exponentially with the hyperparameter $\gamma$. 

\begin{theorem}\label{th:type-1}
Suppose we have $n$ independent observations $Y_1, \ldots, Y_n$ such that each $Y_i$ has a $\mbox{Poi}(\theta_i)$ distribution, and the $\theta_i$'s are drawn from the two-component mixture distribution in \eqref{eq:twogroups}. Under the assumption $\tau \to 0$ as $n \to \infty$, the type I  error probability is given by: 
$$
t_{1i} \equiv t_1 = \mbox{pr}_{H_{0i}}\left\{ E(\kappa_i | Y_i, \gamma, \tau) < 1- \xi \right\} \leq \frac{ \bigg( \frac{\beta}{1+\beta} \bigg)^{\gamma + 1/2}\bigg( \frac{1}{1+\beta} \bigg)^{\alpha-1} }{ (\gamma+1/2) B(\gamma+1/2,\alpha)}.
$$
\end{theorem}
Theorem \ref{th:type-1} shows the importance of the additional parameter $\gamma$ in controlling the type I  error probability when the null distribution of $\theta_i$'s is positive and non-degenerate. In particular, the type-1 error rate for {\small GH} prior would be lower than that of the three-parameter beta prior that arises as a special case of {\small GH} prior for $\gamma = 1$, since it has a spike at zero and heavy-tail but no mechanism for flexible thresholding. The proof is deferred to the supplementary section S.1. 

\subsection{Robustness of Posterior Mean}
\cite{efron2011tweedie} introduced the following simple empirical Bayes estimation formula, called Tweedie's formula, for one parameter exponential families. Suppose $f(y_i \mid \eta_i) = f_{\eta_i}(y_i) = e^{\eta_i y_i - \psi(\eta_i)}f_0(y_i)$, $\eta_i \sim g(\eta_i)$, where $\eta_i$ is the canonical parameter of  the family and $f_0(y_i)$ is the density when $\eta_i=0$. Then the posterior of $\eta_i$ given $y_i$ is: 
\beq
g(\eta_i \mid y_i) = e^{\eta_i y_i - \lambda(y_i)}\{ g(\eta_i)e^{-\psi(\eta_i)} \} \; \mbox{where } \lambda(y_i) = \log\bigg\{ \frac{f(y_i)}{f_0(y_i)} \bigg\},
\eeq
with $f(\cdot)$ the marginal density $f(y_i) = \int f_{\eta_i}(y_i) g(\eta_i ) d \eta_i$. The posterior distribution of $\eta_i$ follows a one-parameter exponential family with canonical parameter $y_i$ and cumulant generating function $\lambda(y_i)$. It follows that the posterior mean and variance can be written as:
\beq
E(\eta_i \mid y_i) = \lambda'(y_i) = l'(y_i)-l_0'(y_i); V(\eta_i \mid y_i) = \lambda''(y_i) = l''(y_i) - l_0''(y_i). \label{eq:tweedy}
\eeq 
The Tweedie's formula when applied for the Poisson mean problem yields the following expression for the posterior mean and variance of $\eta_i = \log(\theta_i)$: 
\beq
E(\eta_i \mid y_i ) = \log \Gamma(y_i+1)'+l'(y_i), \; V(\eta_i \mid y_i) = \log \Gamma(y_i+1)''+l''(y_i), \label{eq:tweedpois}
\eeq
where $l(y_i)= \log m(y_i) = \log\{ \int P_{\eta_i}(y_i) dG(\eta_i)\}$ denotes the log-marginal density of $y_i$. The following theorem establishes the tail robustness properties for the canonical parameter $\log(\theta_i)$.
\begin{theorem}\label{theorem:tweedie}
Suppose $y_i \sim \mbox{Poi} (\theta_i)$, $i = 1,\ldots,n$ and let $m(y_i)$ denote the marginal density of $y_i$ under the Gauss Hypergeometric prior for fixed $\tau$ and $\gamma$, $(\theta_i \mid \kappa_i) \sim \mbox{\small Ga}\{ \alpha, \kappa_i(1-\kappa_i)^{-1}\}$ and $(\kappa_i \mid \tau, \gamma) \sim \mbox{\small GH}(1/2,1/2,\gamma,z=\tau^2-1)$. Then, 
\begin{align*}
\lim_{y_i \rightarrow \infty} m'(y_i)/m(y_i) & = 0 \\
| E(\log \theta_i \mid y_i) -\log(y_i) | & \to 0 \mbox{ as } y_i \to \infty. 
\end{align*}
\end{theorem}
The importance of Theorem \ref{theorem:tweedie} lies in proving the `Bayesian robustness' properties of the {\small GH} prior for the canonical parameter $\eta_i = \log \theta_i$, which guarantees against under-estimation of $\eta_i$ for large values of $\log(y_i)$. The conclusions of Theorem \ref{theorem:tweedie} and the super-efficiency property shown in Theorem S.1 will also hold for the horseshoe or three-parameter beta prior with similar asymptotic rates. The {\small GH} prior stands out from the rest due to its flexible shrinkage property in Theorems \ref{theorem:small-tau} and \ref{th:type-1} that enable it to shrink observations relative to the parameter $\gamma$. 

\section{Simulation Studies}\label{sec:app}

\subsection{Sparse Count Data}

In this section, we report two simulation studies to compare the performance of the different estimators for a high dimensional sparse Poisson mean vector.  We compare our {\small GH} estimator  with the Horseshoe estimator, the Kiefer--Wolfowitz nonparametric maximum likelihood estimator, Robbins' frequency ratio estimator, a Bayesian zero-inflated Poisson estimator and a global shrinkage Bayes estimator. The `Horseshoe' prior is given by:
$$
\theta_i \sim \mathrm{Ga}(\alpha, \lambda_i^2 \tau^2),\quad 
\lambda_i \sim C^+(0,1),\quad \tau \sim C^+(0,1),
$$
where $C^+(0,1)$ denotes a standard half-Cauchy distribution. For the Bayesian zero-inflated Poisson model, we use a Gamma hyperprior on the Poisson mean and a Beta prior on the zero-occurence probability, where the hyper-parameters are estimated from the data. We use average Bayes risk $ABR(\theta) =n^{-1}{E}_{\Theta}(\norm{ \hat{\theta}-\theta }^2)$ as the estimation performance criterion. 
The global shrinkage estimator is obtained by putting a standard conjugate gamma prior on the Poisson means. The parameters $\theta_i$ and the observations are drawn from the model:
\begin{align*}
\theta_i & \sim (1-\omega) \delta_{\{0\}} +\omega |t_{ 3}|, \mbox{ and } Y_i \sim \mbox{Poi}(\theta_i), \; i = 1, \ldots, n, 
\end{align*}
with $|t_{3}|$ denoting a folded t-distribution with 3 degrees of freedom. We generate $1,000$ different datasets from the above model for each combination of multiplicity $n = 200, 500$ and proportion of non-zero parameters $\omega = 0{\cdot}1, 0{\cdot}15, 0{\cdot}2$. For each of the datasets, we estimate $\theta$ and report the mean and standard deviations for the squared error loss for each of the estimators based on these $1,000$ simulations. 

\begin{figure}[!ht]
\begin{subfigure}[t]{0.5\linewidth}
  \centering
  \includegraphics[width=\linewidth]{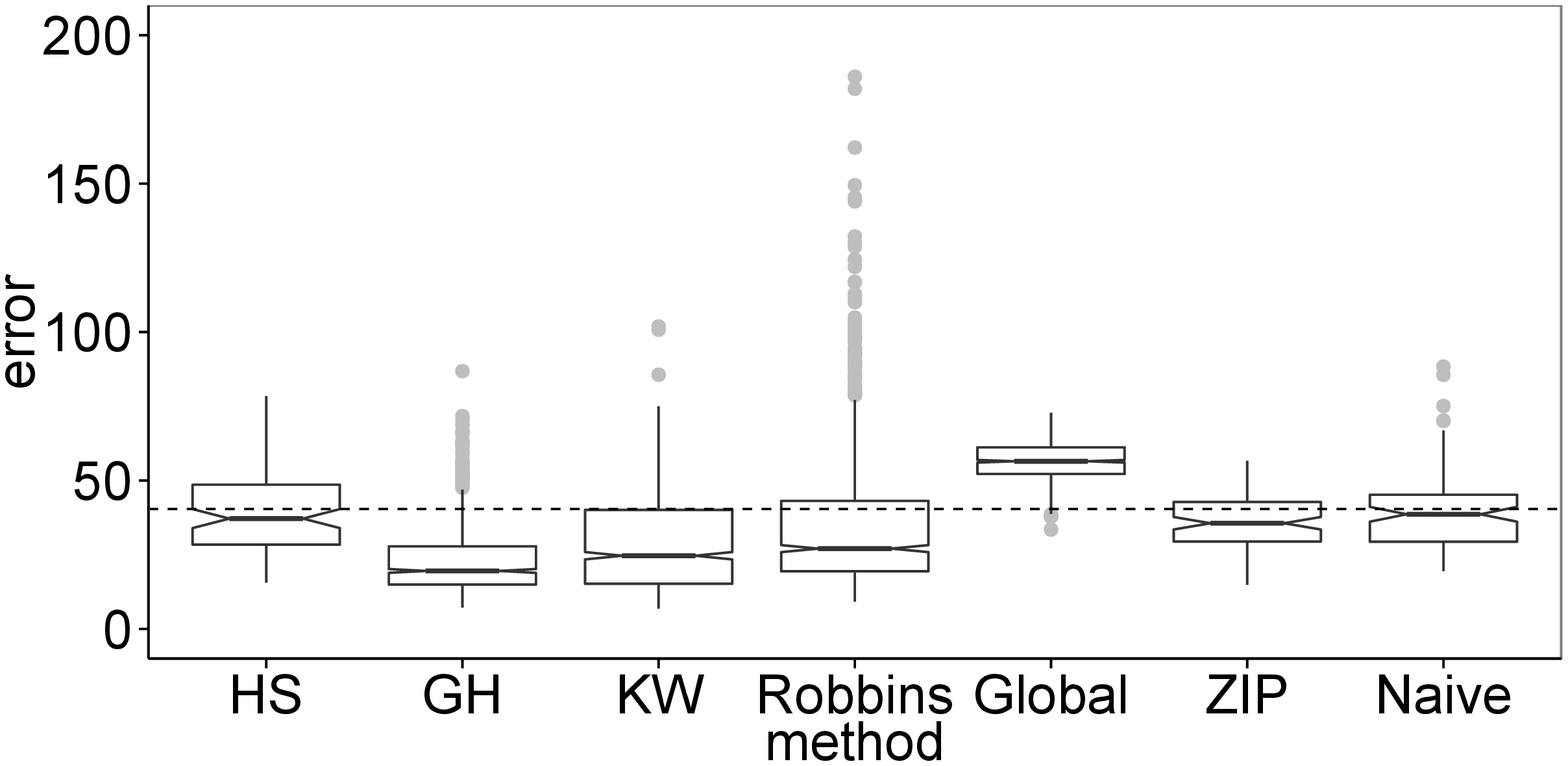}
  \caption{$\omega = 0{\cdot}2$}
  \label{fig:sub1}
\end{subfigure}
\begin{subfigure}[t]{0.5\linewidth}
  \centering
  \includegraphics[width=\linewidth]{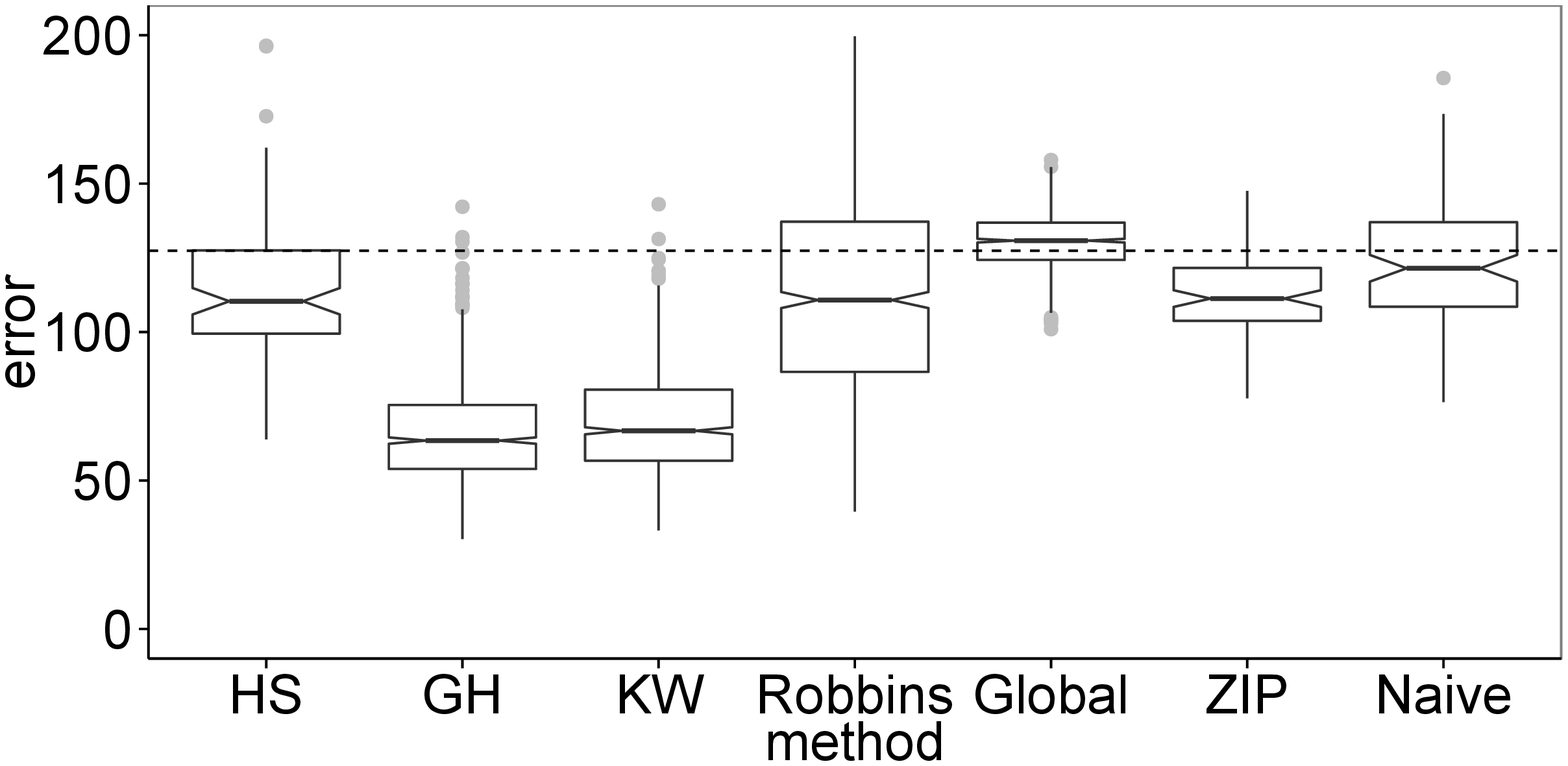}
  \caption{$\omega = 0{\cdot}2$}
  \label{fig:sub2}
\end{subfigure}
\caption{Boxplots for the estimation errors of the competing estimators, namely Horseshoe (HS), Gauss Hypergeometric (GH), Kiefer--Wolfowitz (KW), Robbins' and Global shrinkage estimator, for $n = 200$ and $n = 500$ for a fixed value of $\omega = 0{\cdot}1$.}
\label{fig:sqerr-1}
\end{figure}

The results are given in Table \ref{tab:sqerr} with the boldfaced entries marking the row winners. Figure \ref{fig:sqerr-1} shows box plots for $\omega = 0{\cdot}2$, and $n = 200$ or $n=500$. As expected, the {\small GH} estimator beats its competitors in this sparse simulation set-up across different values of multiplicity and sparsity. The Kiefer--Wolfowitz method also performs well, and is a close runner-up in terms of accuracy, while the  Bayesian ZIP model comes third, with better performance for sparser situations. The frequency ratio estimator does poorly, and the difference is more prominent for higher multiplicity. The global shrinkage prior has the lowest accuracy, as it is not built to handle large sparse mean vectors. We also report the na\"{\i}ve risk $(1/n)\hat{E}(\vectornorm{y-\theta_0}^2)$ in both Table \ref{tab:sqerr} and Figure \ref{fig:sqerr-1} as a baseline to highlight the poor performance of the Robbins' estimator for sparse data. 

\begin{table}
 \centering
\def~{\hphantom{0}}
\caption{Comparison of the average Bayes risk $ABR(\theta) = n^{-1} {E}_{\Theta}(\norm{ \hat{\theta}-\theta }^2)$ for different Bayes and empirical Bayes procedures for $\theta_i \sim (1-\omega) \delta_{\{0\}} +\omega |t_{ 3}|, i = 1, \ldots, n.$}{%
\footnotesize
\begin{tabular}{rrrrrrrrr}
      & Methods & HS    & KW    & GH    & Robbins & Global & ZIP   & Na\"{\i}ve \\
$n = 200$ & $w = 0{\cdot}1$ & $10.2 (3.7)$ & $11.5 (7.1)$ & \boldmath{}\textbf{$8.2 (6.1)$}\unboldmath{} & $15.6 (11.4)$ & $29.6 (2.7)$ & $11.2 (3.3)$ & $12.5 (6.0)$ \\
      & $w = 0{\cdot}15$ & $14.9 (6.3)$ & $6.3 (2.5)$ & \boldmath{}\textbf{$5.5 (2.8)$}\unboldmath{} & $14.1 (10.0)$ & $9.4 (0.8)$ & $8.4 (2.2)$ & $12.7 (4.7)$ \\
      & $w = 0{\cdot}2$ & $19.6 (7.1)$ & $14.0 (7.5)$ & \boldmath{}\textbf{$11.4 (5.9)$}\unboldmath{} & $17.8 (12.4 )$ & $28.1 (3.3)$ & $17.9 (4.5)$ & $19.9 (7.1)$ \\
      &       &       &       &       &       &       &       &  \\
$n = 500$ & $w = 0{\cdot}1$ & $11.1 (3.4)$ & $8.7 (3.2)$ & \boldmath{}\textbf{$7.4 (3.1)$}\unboldmath{} & $16.6 (9.5)$ & $20.2 (1.2)$ & \boldmath{}\textbf{$7.4 (1.5)$}\unboldmath{} & $9.2 (2.4)$ \\
      & $w = 0{\cdot}15$ & $18.6 (4.7)$ & \boldmath{}\textbf{$8.0 (1.8)$}\unboldmath{} & $8.8 (2.5)$ & $18.4 (10.7)$ & $15.2 (0.9)$ & $12.7 (1.9)$ & $16.3 (4.4)$ \\
      & $w = 0{\cdot}2$ & $22.8 (5.5)$ & $13.8 (3.3)$ & \boldmath{}\textbf{$13.1 (3.6)$}\unboldmath{} & $27.4 (14.9 )$ & $26.1 (1.9)$ & $22.5 (2.8)$ & $24.8 (4.4)$ 
\end{tabular}%
}
\label{tab:sqerr}
\end{table}

\subsection{Multiple Testing}\label{sec:mtp}

As argued in \S \ref{sec:t1}, thresholding the shrinkage weights $(1-\hat{\kappa_i})$ under the {\small GH} prior induces a multiple testing rule \citep{polson2012good}.  For the Kiefer-Wolfowitz estimator, we  threshold the shrinkage weights $\omega_i = \hat{P}_G(y_i)^{-1} \hat{P}_G(y_i+1)$, where $\hat{P}_G(\cdot)$ is the estimated probability mass function. We show below that the {\small GH} decision rule dominates that induced by the three parameter beta prior \citep{armagan2011generalized} and the Kiefer--Wolfowitz estimator. We choose $a = b = 1/2$ and $ z= \tau^2-1$ as the hyper-parameter values for the three parameter beta. 

\begin{figure}[!ht]%
\centering
\includegraphics[width=0.8\linewidth]{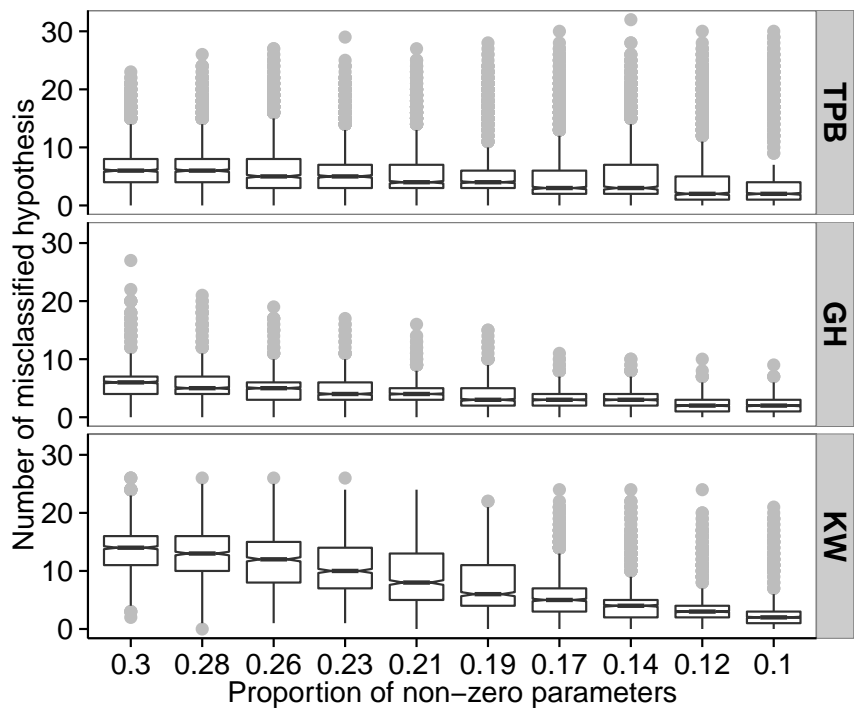}%
\caption{Number of misclassified hypotheses under competing decision rules, the three-parameter beta (TPB), Gauss Hypergeometric (GH), and the Kiefer--Wolfowitz (KW) for different values of the proportion of non-null effects. }
\label{fig:misclass}%
\end{figure}

We generate $n = 200$ observations from a `contaminated' zero-inflated model: $y_i$ is either zero with probability $1-\omega$ or drawn from a $\mbox{Poi}(4)$ with probability $\omega$. We contaminated the data by setting $p$ proportion of the $0$'s equal to $1$. Our goal is to detect the non-null parameters. The non-null parameter value $\lambda= 4$ is chosen to be of the order of the maximum order statistics $M_n$ of $n$ independent and identically distributed Poisson random variables.  It can be shown that $M_n$ lies inside the interval $[I_n, I_n+1]$ with probability $1$ as $n \to \infty$, where $I_n \sim \log n / \log \log n$ \citep{anderson1970extreme}, and the midpoint of the interval is $I_{200} + 0 {\cdot} 5= 3{\cdot}67$. Following \S \ref{sec:t1}, the decision rule induced by the shrinkage priors is to reject the $i^{th}$ null hypothesis if $1-E(\kappa_i | y_i, \tau, \gamma) > \xi$ for some fixed threshold $\xi$. We calculated this threshold by applying the \textit{kmeans} clustering algorithm on the shrinkage weights with number of clusters $k = 2$ and setting $\xi$ to be mean of the cluster centers. To compare the shrinkage priors, we calculate the number of misclassified hypotheses for $1,000$ simulated datasets for $10$ equidistant values of the proportion of `sparsity' $\omega \in [0{\cdot}1,0{\cdot}3]$. We fixed the value of the contamination proportion $p = 0{\cdot}1$. The box plots for the number of misclassification errors, shown in Figure \ref{fig:misclass}, and the mean number of misclassified hypothesis, given in Table \ref{tab:misclass}, suggest that the {\small GH} decision rule outperforms both the three-parameter beta and the Kiefer--Wolfowitz method for all sparse scenarios. Further simulation studies in supplementary section S.3 show that the {\small GH} prior beats its competitors when the $\theta_i$'s are drawn from a mixture of Gamma distributions with the null favoring higher concentration near zero. 

\begin{table}[ht]
\centering
\def~{\hphantom{0}}
\caption{Mean misclassification errors for the three parameter beta prior, the Gauss Hypergeometric prior and the Kiefer--Wolfowitz non-parametric maximum likelihood estimator.}{%
\footnotesize
     \begin{tabular}{lcccccccccc}
    Sparsity & $0{\cdot}3$ & $0{\cdot}27$ & $0{\cdot}25$ & $0{\cdot}23$ & $0{\cdot}21$ & $0{\cdot}18$ & $0{\cdot}16$ & $0{\cdot}14$ & $0{\cdot}12$ & $0{\cdot}1$ \\
		\\
    TPB    & $7{\cdot}1$ & $7{\cdot}1$ & $7{\cdot}0$ & $6{\cdot}5$ & $6{\cdot}5$ & $6{\cdot}3$ & $6{\cdot}2$ & $6{\cdot}3$ & $5{\cdot}8$ & $5{\cdot}6$ \\
    GH    & $\mathbf{5{\cdot}9}$ & $\mathbf{5{\cdot}4}$ & $\mathbf{4{\cdot}7}$ & $\mathbf{4{\cdot}4}$ & $\mathbf{4{\cdot}0}$ & $\mathbf{3{\cdot}5}$ & $\mathbf{3{\cdot}0}$ & $\mathbf{2{\cdot}7}$ & $\mathbf{2{\cdot}1}$ & $\mathbf{1{\cdot}7}$ \\
    KW    & $13{\cdot}7$ & $12{\cdot}7$ & $11{\cdot}7$ & $10{\cdot}5$ & $9{\cdot}0$ & $7{\cdot}6$ & $6{\cdot}1$ & $4{\cdot}8$ & $3{\cdot}4$ & $2{\cdot}8$     
    \end{tabular}}%
  \label{tab:misclass}%
\end{table}

\section{Detecting Rare Mutational Hotspots}\label{sec:wgs}
In this section, we apply our methods to count data arising from a massive sequencing study called the Exome aggregation consortium (ExAC). The ExAC database reports the total number of mutated alleles or variants along the whole exome for 60,076 individuals, and provides information about genetic variation in the human population. It is widely recognized in the scientific community that these rare changes are responsible for both common and rare diseases \citep{pritchard2001rare}. The frequency of mutated alleles is very low or zero for a vast majority of the locations across the genome and substantially higher in certain functionally relevant genomic locations, such as promoters and insulators \citep{lewin2011lewin}. An important problem in rare mutations is identification of such mutational `hotspots' where the mutation rate significantly exceeds the background. This is important since these mutational hotspots might be enriched with disease-causing rare variants \citep{ionita2012scan}. 

Our goal is to use the shrinkage prior developed in this paper to identify the potential hotspots harboring rare variants in a genomic region. Towards this, we first filter out the `common' variants with minor allele frequency greater than $0{\cdot}05\%$, a threshold suggested by \citet{cirulli2015exome}, on a gene PIK3CA known to be responsible for ovarian and cervical cancers \citep{shayesteh1999pik3ca}. The mutation dataset contains the number of `mutated' alleles along the gene PIK3CA for $240$ amino acid positions ranging from $0$ to $1066$. Intuitively, the flexible shrinkage property of the {\small GH} prior will lead to a better detection of the `true' hotspots by better shrinking low counts. The number of mutated alleles at the $i^{th}$ position, denoted by $Y_i$, ranges from $0$ to $58$. 
We model $y_i \sim \mbox{Poi}(N_i \theta_i)$ independently, where $\theta_i$ is the mutation rate and $N_i$ is the number of alleles sequenced at location $i$. Here we make the simplifying assumption of uniform sequencing depth across the gene such that $N_i = N$ for all $i$, but in general the sequencing depth is dependent on location. Since each individual carries two copies of alleles that could harbor a rare variant, $N_i$ is also equal to twice the number of individuals sequenced at that position.
 
\begin{figure}[!ht]
\centering
\includegraphics[width=0.75\linewidth]{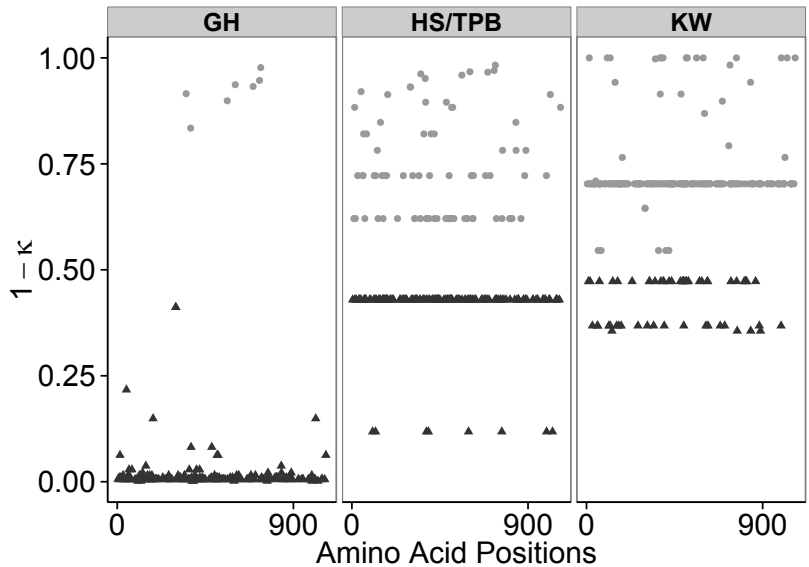}%
\caption{The histogram of the count data and a comparison of the shrinkage profile for different estimators, namely, Gauss Hypergeometric (GH), three-parameter beta (TPB) / Horseshoe (HS) prior, and the Kiefer--Wolfowitz (KW) nonparametric maximum likelihood estimator.}
\label{fig:real}%
\end{figure}
We compare  the shrinkage profile of the {\small GH} prior with the three-parameter beta / horseshoe prior and the Kiefer--Wolfowitz  estimator in terms of the number of identified mutational hotspots. We apply the multiple testing rule proposed in \S \ref{sec:mtp} to identify the variants for which the number of mutated alleles are substantially higher than the background. The number of variants identified as `non-null' by the three methods, namely, {\small GH} prior, three-parameter Beta/horseshoe prior and the Kiefer--Wolfowitz method are $7$, $81$ and $56$ respectively. Figure \ref{fig:real} shows the pseudo inclusion probabilities of the competing methods and clearly demonstrates that the {\small GH} prior has a sharpened ability of segregating the substantially higher signals from the background noise. 
\section{Application to Global Terrorism Data}\label{sec:gtd}

\begin{figure}[!ht]%
\centering
\includegraphics[width=0.75\linewidth]{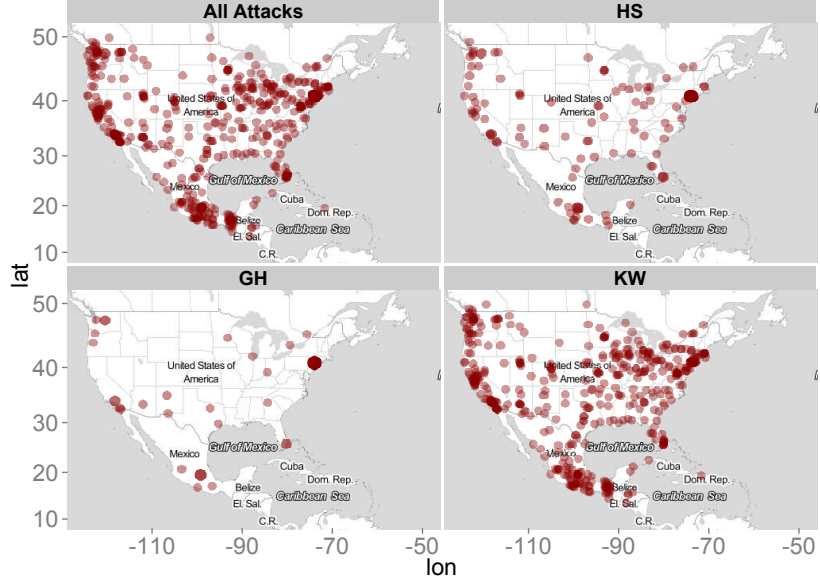}%
\caption{The total number of terror attacks in North American cities since $1970$ and the posterior mean estimates of the rate of the terror attacks by three different methods, the Horseshoe (HS), the Gauss-Hypergeometric (GH), and the Kiefer--Wolfowitz (KW). The size and the opacity of the points on the graph increases with the value of the observation plotted.}%
\label{fig:gtd}%
\end{figure}

We consider an application to the global terrorism database containing details about all terrorist attacks in the world since 1970 along with the location and type of such attacks. We focus on terrorist attacks in North America, aggregated over the years of observation at the level of cities, and aim to find the cities that are worst hit. As expected, there are many cities with zero and small counts with a few large counts like New York, Mexico City, Miami etc. Our goal is to apply the {\small GH} prior to select the cities with high attack rates.  Figure \ref{fig:gtd} shows the observed counts of terrorist attacks along with the posterior mean estimates for the rates for the North American cities for the three different methods under consideration, Horseshoe, Gauss-Hypergeometric and Kiefer--Wolfowitz. Since there are many cities with exact zero counts, we only show here the non-zero observations.  As expected, the {\small GH} method selects the fewest cites (25) while the Kiefer--Wolfowitz selects every city having at least one attack.
This illustrates the tendency of {\small GH} to be much more robust to very small counts that may correspond to very sporadic  events or errors in recording; for example, an attack may be mislabeled as terrorism-related.  Methods that naively select all locations with non-zero counts are not very useful in practice.  

\section*{Acknowledgment}
The authors thank Dr. Sandeep Dave for invaluable help in learning about exome sequencing data.  This research was funded by the National Science Foundation through the Statistical and Applied Mathematical Sciences Institute and a grant from the National Institutes of Health.  We also thank the reviewers for excellent comments that led to substantial improvements in the article.

\section*{Supplementary material}
\label{SM}
The supplementary file contains theoretical results for Kullback-Leibler super-efficiency, a proof for Theorem \ref{th:type-1}, and its validation through a simulation study, more details about the data-set used in \S5, and performance of the {\small GH} prior for zero-inflated data without covariates. 
\section*{Appendix 1}
\subsection*{Proof of Lemma \ref{lemma:1}}
\begin{proof}
Using Laplace transformation $1/(c+z)=\int_0^\infty \mathrm{e}^{-(c+z)t} dt$ gives the following:
\begin{align*}
E\{1/(c+Z)\} & =\int_0^\infty\mathrm e^{-ct}\int_{0}^{\infty}\frac{\beta^{\alpha}}{\Gamma(\alpha)} z^{\alpha-1}\mathrm e^{-(\beta+t) z} \mathrm{d} z \mathrm{d}t =\int_0^\infty\mathrm e^{-ct}\frac{\beta^{\alpha}}{(\beta+t)^{\alpha}}\mathrm dt \\
& = \mathrm{e}^{c\beta}\beta^{\alpha}c^{1-\alpha}\int_{c\beta}^\infty s^{-\alpha}\mathrm e^{-s} \mathrm{d}s = \mathrm{ e}^{c\beta}\beta^{\alpha}c^{1-\alpha}\Gamma(1-\alpha, \beta c).
\end{align*}
where the last step follows by changing variable $s=c(\beta+t)$.
\end{proof}
\subsection*{Proof of Proposition \ref{prop:marginal}}
\begin{proof}
For simplicity of notation, we drop the subscript $i$ in the proofs and use $\theta$, $y$ and $\kappa$ to denote a generic element of the corresponding vector rather than the vector itself. Transforming $ u = 1/\lambda^2$, we get:
\begin{align*}
p(\theta) & =  \frac{2}{\pi \Gamma(\alpha) } \int_{0}^{\infty} {\theta}^{\alpha-1} \exp(-\theta/\lambda^2) {\lambda}^{-2\alpha}  \frac{1}{\lambda^2+1} d\lambda  \\
& = \frac{1}{\pi \Gamma(\alpha) } \int_{0}^{\infty} \theta^{\alpha-1} \exp(-\theta u) u^{\alpha - 1/2} \frac{1}{1+u} du \\
& = \frac{\Gamma( \alpha + 1/2) }{\pi \Gamma(\alpha) } \theta^{-3/2} E \{ 1/(1+U) \} \quad \mbox{ where } U \sim \mathrm{Ga}\bigg(\alpha+1/2, \theta \bigg) \\
& =  \frac{1}{\sqrt{\pi} \beta(\alpha,1/2) } \mathrm{e}^{\theta} \theta^{\alpha-1} \Gamma \left(1/2 - \alpha,\theta \right).
\end{align*}
where the last step follows from Lemma \ref{lemma:1} with $c = 1$.
\end{proof}

\subsection*{Proof of Corollary \ref{cor:marginal}}
\begin{proof}
For the first set of bounds for $\alpha = 1$, we use the recurrence relation for incomplete gamma functions $\Gamma(x+1,y) = x\Gamma(x,y)+y^{x} e^{-y}$ to derive the following: 
\begin{align*}
p(\theta \mid \alpha = 1) & = \frac{1}{2\sqrt{\pi}} e^{\theta} \Gamma(-1/2,\theta) = \frac{1}{\sqrt{\pi}} e^{\theta} (\theta^{-1/2}e^{-\theta}-\Gamma(1/2,\theta)) \\
& = \frac{1}{\sqrt{\pi}}\theta^{-1/2}-e^{\theta}\mathrm{erfc}(\sqrt{\theta})) 
\end{align*}
The bounds follow from the following inequality \citep{abramowitz1964handbook}, 7.1.13: 
\begin{equation*}
\left( x+\sqrt{x^2+2} \right)^{-1}  \leq e^{x^2} \int_{x}^{\infty} e^{-t^2} dt \leq \left( x+\sqrt{x^2+4/\pi} \right)^{-1}
\end{equation*}
For $\alpha = 1/2$, the prior density $p(\theta)$ can be written as: 
\begin{equation}
p(\theta \mid \alpha = 1/2) = \frac{1}{\sqrt{\pi} \beta(1/2,1/2) } \mathrm{e}^{\theta} \theta^{-1/2} \Gamma \left(0,\theta \right)  = \frac{1}{\pi^{3/2} } \mathrm{e}^{\theta} \theta^{-1/2} E_1(\theta) \label{eq:alpha-half}
\end{equation}
where $E_1({\cdot})$ is the exponential integral function defined as $ E_1(z) = \int_{z}^{\infty} e^{-t} /t dt$, and satisfies the following tight upper and lower bounds: 
\begin{align*}
(1/2) e^{-t} \log( 1+ 2/t) < E_1(t) < e^{-t} \log(1+1/t) 
\end{align*}
The bounds on the prior density for $\alpha = 1/2$ can be easily derived by using these bounds on the exponential integral and the expression for the marginal prior in Equation \eqref{eq:alpha-half}. 
\end{proof}

\subsection*{Proof of Theorem \ref{theorem:large-y}} 
\begin{proof}
To prove the theorem, we will use the fact that the posterior density of $\kappa$ can be written as the product of $p_1(\kappa) = \sqrt{\kappa}, p_2(\kappa) = (1-\kappa)^{y-1/2}$ and $p_3(\kappa)=(1-(1-\tau^2)\kappa)^{-\gamma}$, and for any $\tau^2 \in (0,1)$ and $\gamma$, $(1-\kappa)^{-\gamma} > (1-(1-\tau^2)\kappa)^{-\gamma}$, which implies 
\begin{equation}
(1-(1-\tau^2)\kappa)^{y -1/2-\gamma} \leq p_2(\kappa) p_3(\kappa) \leq (1-\kappa)^{y-1/2-\gamma} \mbox{ for } \gamma \geq 0, \tau^2 \geq 0 \label{eq:order}
\end{equation} 
Clearly, both the bounds in Equation \ref{eq:order} are increasing functions of $\kappa \in (0,1)$ for any $y > 1/2+ \gamma$ and can be trivially bounded by its values at the boundaries. We derive the following upper bound on the tail probability of the posterior distribution of $\kappa$ given $y$, $\tau$, and $\gamma$. 
\begin{align*}
P(\kappa > \eta \mid y, \tau, \gamma) & \leq \frac{\int_{\eta}^{1} \kappa^{1/2} (1-\kappa)^{y-1/2} (1-(1-\tau^2)\kappa)^{-\gamma} d\kappa}{\int_0^{\eta} \kappa^{1/2} (1-\kappa)^{y-1/2} (1-(1-\tau^2)\kappa)^{-\gamma} d\kappa} \\
&  \leq \frac{\int_{\eta}^{1} \kappa^{1/2} (1-\kappa)^{y-1/2-\gamma} d\kappa}{(1-\tau^2)^{-1/2} \int_0^{\eta (1-\tau^2) } z^{1/2} (1-z)^{y-1/2-\gamma} dz} \\
& \leq \frac{(1-\eta)^{y-1/2-\gamma} (1-\eta^{3/2} ) } {(1-(1-\tau^2)\eta)^{y-1/2-\gamma} (1-\tau^2) \eta^{3/2} } = C(\eta)\frac{1}{1-\tau^2} \left(1 + \frac{\tau^2\eta}{1-\eta} \right)^{-(y-1/2-\gamma)}
\end{align*}
The proof follows from the fact that the right hand side of the last equation goes to zero as $y$ goes to infinity for any non-zero $\eta$ and $\tau^2$. 
\end{proof}

\section*{Appendix 2}
\subsection*{Proof of Theorem \ref{theorem:small-tau}} 
\begin{proof}
We use the same factorization as in the proof of \ref{theorem:large-y} and the fact that both the bounds in \ref{eq:order} are decreasing functions of $\kappa$ for $y \leq  \gamma - 1/2$. Also, we denote $y-(\gamma-1/2) = -d < 0$ for small $y$. 
\begin{align*}
P(\kappa < \eta \mid y, \tau, \gamma) & \leq \frac{\int_{0}^{\eta} \kappa^{1/2} (1-\kappa)^{-d-1} d\kappa}{\int_{\eta}^{1} \kappa^{1/2} (1-(1-\tau^2)\kappa)^{-d-1} d\kappa} \leq \frac{\sqrt{\eta} \int_{0}^{\eta} (1-\kappa)^{-d-1} d\kappa}{\sqrt{\eta} (1-\tau^2)^{-1} \int_{\tau^2}^{1-(1-\tau^2)\eta} (z)^{-d-1} dz} \\
& \leq \frac{ (1-\tau^2) ((1-\eta)^{-d} -1) }{ \tau^{-2d} - (1-(1-\tau^2)\eta)^{-d} } = \frac{ (1-\tau^2) ((1-\eta)^{-d} -1) }{ \tau^{-2d}(1+o(1))} \leq \left( \frac{\tau^2}{1-\eta} \right)^{d}
\end{align*}
Hence for any $\eta \in (0,1)$, the posterior mass of $\kappa$ in the interval $(0,\eta)$ will shrink to zero for all values of $y \in (0,\gamma-1/2)$ as $\tau \to 0$. 
\end{proof}
\begin{proof}
For the first part of Theorem \ref{theorem:tweedie}, we note the following: 
\begin{align*}
\frac{m'(y)}{m(y)} & = \frac{\frac{d}{dy} \int_0^1 \kappa^{1/2} (1-\kappa)^{y-1/2} (1-(1-\tau^2)\kappa)^{-\gamma} d\kappa}{\int_0^1 \kappa^{1/2} (1-\kappa)^{y-1/2} (1-(1-\tau^2)\kappa)^{-\gamma} d\kappa} \\
& = \frac{\int_0^1 \kappa^{1/2} (\frac{d}{dy}(1-\kappa)^{y-1/2}) (1-(1-\tau^2)\kappa)^{-\gamma} d\kappa}{\int_0^1 \kappa^{1/2} (1-\kappa)^{y-1/2} (1-(1-\tau^2)\kappa)^{-\gamma} d\kappa} = \frac{1}{\log(y-1/2)} 
\end{align*}
Hence from Tweedie's formula for Poisson mean (\textit{vide} Equation \ref{eq:tweedpois}), it follows that:
\begin{equation}
\mid E(\eta \mid y) - \log(y)\mid = (\psi(y+1)-\log(y)) + {\log(y-1/2)}^{-1} \label{eq:th1.1}
\end{equation}
where $\psi(z) = \frac{d}{dz}\log \Gamma(z)$ is the Digamma function \citep{abramowitz1964handbook}, Ch. 6 that satisfies: 
\beq
\psi(z) \asymp \log(z) -\frac{1}{2z}-\frac{1}{12z^2} + \frac{1}{120z^4} + \dots = \log(z) + O \left(\frac{1}{z} \right) \label{eq:th1.2}
\eeq
Using Equations \eqref{eq:th1.1} and \eqref{eq:th1.2} together, we arrive at 
\beq
\mid E(\eta \mid y) - \log(y) \mid = \log (1+y^{-1}) + {\log(y-1/2)}^{-1} +  O\left( y^{-1} \right)  = O\left( {\log(y)}^{-1} \right) \nonumber
\eeq
where the right hand side of the above equation goes to zero as $y \to \infty$. 
\end{proof}

\bibliographystyle{apalike}
\bibliography{poissonref}

\end{document}

%% file: math-commands.tex
\renewcommand{\S}{{\bf S}}

\newcommand{\ben}{\begin{enumerate}}
\newcommand{\een}{\end{enumerate}}
\newcommand{\beq}{\begin{equation}}
\newcommand{\eeq}{\end{equation}}

\newcommand{\norm}[1]{\lVert#1\rVert}
\newcommand{\vectornorm}[1]{\left|\left|#1\right|\right|}

\newtheorem{theorem}{Theorem}
\numberwithin{theorem}{section}

\numberwithin{Def}{section}
\newtheorem{remark}{Remark}
\numberwithin{remark}{section}

\newtheorem{proposition}{Proposition}
\numberwithin{proposition}{section}
\newtheorem{lemma}{Lemma}
\numberwithin{lemma}{section}
\newtheorem{Cor}{Corollary}
\numberwithin{Cor}{section}